\documentclass[aip,jap,floatfix,amsmath,amssymb,reprint]{revtex4-1}

\usepackage{newtxtext,newtxmath}
\usepackage[T1]{fontenc}
\usepackage{dcolumn}
\usepackage{graphicx}
\usepackage{booktabs}
\usepackage{etoolbox}
\usepackage[dvipsnames]{xcolor}
\usepackage{floatrow}
\usepackage{multirow}

\usepackage[pdftex]{hyperref}
\hypersetup{
  pdftitle={MCFETmodel},
	colorlinks,
	citecolor=blue,
	linkcolor=blue
	}

\draft 

\makeatletter
\def\@email#1#2{%
 \endgroup
 \patchcmd{\titleblock@produce}
  {\frontmatter@RRAPformat}
  {\frontmatter@RRAPformat{\produce@RRAP{*#1\href{mailto:#2}{#2}}}\frontmatter@RRAPformat}
  {}{}
}%
\makeatother

\begin{document}

\title{
A Compact Model for Polar Multiple-Channel Field Effect Transistors: A Case Study in III-V Nitride Semiconductors}

\author{Aias~Asteris}
\email[Author to whom correspondence should be addressed: ]{aa2484@cornell.edu, djena@cornell.edu}
\affiliation{\hbox{Department of Materials Science and Engineering, Cornell University, Ithaca, New York 14853, USA}}
\author{Thai-Son~Nguyen}
\affiliation{\hbox{Department of Materials Science and Engineering, Cornell University, Ithaca, New York 14853, USA}}
\author{Huili G.~Xing}
\affiliation{\hbox{Department of Materials Science and Engineering, Cornell University, Ithaca, New York 14853, USA}}
\affiliation{\hbox{School of Electrical and Computer Engineering, Cornell University, Ithaca, New York 14853, USA}}
\affiliation{\hbox{Kavli Institute at Cornell for Nanoscale Science, Cornell University, Ithaca, New York 14853, USA}}

\author{Debdeep~Jena}
\affiliation{\hbox{Department of Materials Science and Engineering, Cornell University, Ithaca, New York 14853, USA}}
\affiliation{\hbox{School of Electrical and Computer Engineering, Cornell University, Ithaca, New York 14853, USA}}
\affiliation{\hbox{Kavli Institute at Cornell for Nanoscale Science, Cornell University, Ithaca, New York 14853, USA}}

\date{\today}

\begin{abstract}
A compact analytical model is developed for the mobile charge density of polar multiple channel field effect transistors. 
Two dimensional electron and hole gases can be potentially induced by spontaneous and piezoelectric polarization in polar heterostructures.
Focusing on the active region of devices that employ a multiple quantum-well layout, the total electron and hole populations are estimated from fundamental electrostatic and quantum mechanical principles.
Hole gas depletion techniques, revolving around intentional donor doping, are modeled and evaluated, culminating in a generalized closed-form equation for the mobile carrier density across the doping schemes examined.
The utility of this model is illustrated for the III-Nitride material system, exploring AlGaN/GaN, AlInN/GaN and AlScN/GaN heterostructures.
The compact framework provided herein considerably elucidates and enhances the efficiency of multi-layered transistor design.
\end{abstract}


\maketitle 

\section{Introduction}
Multilayered heterostructures based on III-Nitride semiconductors (GaN, AlN, InN etc.)~have gained significant attention due to their superior material properties, including highly-tunable direct bandgaps, high electron mobility, and strong polarization effects.
Multilayered III-Nitride structures have enabled advanced photonic applications, such as laser diodes\cite{Nakamura_1996,Nakamura1998} and distributed Bragg reflectors\cite{Khan1991,Ng1999,vanDeurzen2023}, where their ability to support high-efficiency light generation and detection is leveraged.
Their presence has been further established in electronic devices, such as multichannel field effect transistors\cite{Gaska1999,R.Howell_2014, BridgeFET,FlorinUdrea2022,TSNguyen2025} (MCFETs), polarization super junctions\cite{FlorinUdreaAnalytical}, and lateral Schottky diodes\cite{Terano_2015, Yuhao2021}, wherein they benefit from the presence of multiple two-dimensional carrier gases to overcome the intrinsic trade-off between charge density and mobility of single channel field effect transistors.
By doing so, multichannel devices exhibit low specific ON-resistance, deliver high output power density, and achieve high energy efficiency.

The performance of the aforementioned devices is fundamentally governed by the distribution and control of mobile carriers. 
The distribution of electrons or holes directly impacts current transport, optical emission, and overall device reliability. 
The ability to predict and manipulate carrier populations is thus a critical point of control for device design and optimization.
Despite this importance, there are limited works pertaining to said systems. 
Instead, device modeling conventionally depends on computationally-expensive numerical calculations, which inherently provide limited understanding of the underlying physics.
Several works analytically investigate two channel nitride FETs \cite{Wei2018,Rahman2019,Malik2021}, while expansion to periodic structures has been typically limited to non-polar materials\cite{Nawaz1996,Fujihira_1997}.
Chen et al.\cite{ChenModel1} have developed a comprehensive model for the two-dimensional electron gas (2DEG) density in polar MCFETs but avoided dealing with the potential formation of a two-dimensional hole gas (2DHG), while He et al.\cite{FlorinUdreaAnalytical} have constructed analytical expressions for the breakdown voltage and figure of merit of similar devices.

In this work, we present a compact analytical model for the density of two-dimensional electron and hole gases in multilayered polar heterostructures to provide a tractable, efficient and physically insightful framework to complement and guide numerical simulations.
Quantum mechanical principles are integrated with electrostatic fundamentals to analytically model the formation of two dimensional carrier gases in the active region of these systems.
For completeness, the first section of this work focuses on single channel structures.
From there, we expand to the MCFET, and eventually model and evaluate hole gas depletion techniques via intentional compensation, by either $\delta$-doping or modulation doping. 
The derivation further allows for the implementation of spacer layers between the barrier and the channel layers.
Interlayers are routinely used to boost electron mobility by suppressing alloy and interface roughness scattering\cite{Kaun_2014,Hardy2017,Sohi_2021}.
The method is illustrated by examining assorted schemes of nitride heterostructures, focusing on GaN-based devices, with AlGaN, AlInN, or AlScN barriers, and (GaN/AlN) interlayers.

\section{Single Channel}

\begin{figure}[ht!]
    \includegraphics[width=1\linewidth]{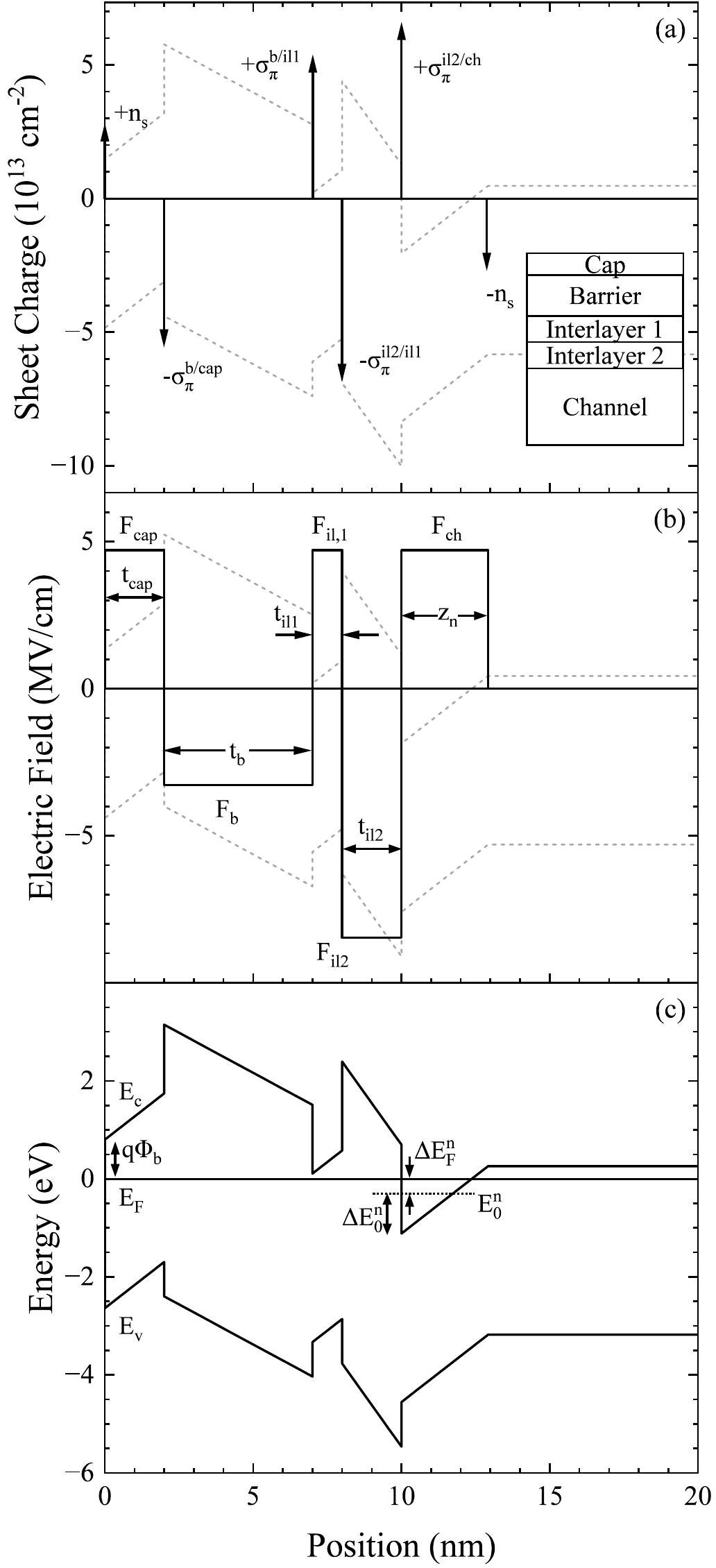}
    \caption{(a) Charge, (b) electric field, and (c) energy band diagrams of an indicative single channel FET. In (a) and (b), the energy band diagram is overlaid (dashed line) as a guide to the eye. All values shown are extracted from the analytical model. The inset plot is a schematic of the epitaxial structure modeled. Notations are explained in the text.}
    \label{SC-CFB diagram}
\end{figure}

Before developing the compact model for the multichannel structure, it is worth visiting the single channel field effect transistor.
This will simplify modeling of the MCFET and further allow for the comparison of the two systems.
There are several works catering to single channel systems, many of which expand to out-of-equilibrium scenarios to map transfer and output characteristics\cite{Ridley2004,Jena2022}.
For this analysis, we will limit ourselves to the equilibrium case.

The single channel model is developed following Fig.~\ref{SC-CFB diagram}, which shows the charge, electric field, and energy band diagrams of a typical single channel FET. 
The 2DEG, denoted by $n_{\rm  s}$ (in cm$^{-2}$), is represented by a sheet charge situated at some distance $z_{\rm n}$ from the respective interface.
$\sigma_\pi^{i/j}$ (in cm$^{-2}$) are the polarization-induced interface sheet charges between layers $i$ and $j$. That is $\sigma_\pi^{i/j} =\frac{1}{q} \left( P_{i} - P_{j}\right),\ i,j = cap,\ b,\ il1,\ il2,\ ch$ where $q$ is the absolute electron charge.
The indices \textit{b}, \textit{il1}, \textit{il2}, and \textit{ch} stand for barrier, first and second interlayer, and channel.
The polarization vectors, $P_i$, include both the spontaneous and the piezoelectric components, and adopt the zinc-blende unit-cell reference\cite{ambacher_2000,Vurgaftman2003,Ambacher2021}.
Moreover, the positive z-axis is defined to run along the [000$\bar1$] crystallographic direction, setting the metal-polar polarization vectors to positive algebraic values.

The consideration of sheet charges dictates that the electric field is constant in between, resulting in the linear variation of energy bands in space.
By extension, each layer $i$ is characterized by its electric field, denoted by $F_i$.
A triangular well forms at the top channel interface, wherein carriers are assumed to be degenerate and to occupy a single subband, $E_0^{\rm n}$.
The position of said subband with respect to the bottom of the conduction band well can be approximated by
\begin{equation}
    \Delta E_0^{\rm n} \approx \left( \frac{9\pi}{8}\right)^{\frac{2}{3}}  \Biggl[ \frac{(qF_{\rm  ch}\hbar)^2}{2m_{\rm c}^{*}}\Biggr]^{\frac{1}{3}}
    \label{ground state},
\end{equation}
where $\hbar$ is the reduced Planck's constant, and $m^{*}_{\rm c}$ is the out-of-plane effective mass for electrons in the conduction band.
The energy difference between the ground state and the Fermi level, $\Delta E_{\rm F}^{\rm n}$, can be further expressed by invoking Fermi-Dirac statistics,
\begin{equation}
        \Delta E_{\rm F}^{\rm n} \equiv E_{\rm F} - E_{\rm 0}^{\rm n} =  k_{\rm  b} T \ln \left({\rm e}^{n_{\rm s}/N_{\rm c}^{\rm 2d}} - 1 \right),
    \label{Fermi}
\end{equation}
where $k_{\rm  b}$ is the Boltzmann constant, $T$ is temperature, and $N_{\rm c}^{\rm 2d} = g_{\rm s} g_{\rm v}\frac{m^{*}_{\rm c}k_{\rm  b} T}{2\pi\hbar^2}$ is the conduction band two-dimensional effective density of states, with $g_{\rm s}$ and $g_{\rm v}$ denoting the spin and valley degeneracy.

The electric field terms are expressed with respect to the 2DEG density, by Poisson's Equation and the displacement vector continuity. 
Starting from the bulk channel region, where the electric field vanishes, we get
\begin{subequations}
\begin{eqnarray}
    F_{\rm ch} &&= \frac{q n_{\rm s}}{\epsilon_{\rm ch}} \label{e-fields1},\\
    F_{\rm il,2} &&= \frac{\epsilon_{\rm ch}}{\epsilon_{\rm il2}} F_{\rm ch} -  \frac{q\sigma_\pi^{\rm il2/ch}}{\epsilon_{\rm il2}} \label{e-fields2},\\
    F_{\rm il,1} &&= \frac{\epsilon_{\rm ch}}{\epsilon_{\rm il1}} F_{\rm ch} -  \frac{q\sigma_\pi^{\rm il1/ch}}{\epsilon_{\rm il1}} \label{e-fields3},\\
    F_{\rm b} &&= \frac{\epsilon_{\rm ch}}{\epsilon_{\rm b}} F_{\rm ch} -  \frac{q\sigma_\pi^{\rm b/ch}}{\epsilon_{\rm b}}, \text{ and}\\
    F_{\rm cap} &&= \frac{\epsilon_{\rm ch}}{\epsilon_{\rm cap}} F_{\rm ch} -  \frac{q\sigma_\pi^{\rm cap/ch}}{\epsilon_{\rm cap}},\label{e-fields4}
\end{eqnarray}
\label{fields_sc}
\end{subequations}
where $\epsilon_i$ is the dielectric constant of layer $i$.

Finally, following the energy band diagram in Fig.~\ref{SC-CFB diagram}(c), we notice that
\begin{equation}
\begin{split}
        q \Phi_{\rm b} + q F_{\rm cap} t_{\rm cap} &-  \Delta E_{\rm c}^{\rm cap/b} + q F_{\rm b} t_{\rm b} \\ - \Delta E_{\rm c}^{\rm b/il1}
         + q &F_{\rm il1} t_{\rm il1} - \Delta E_{\rm c}^{\rm il1/il2}
         + q F_{\rm il2} t_{\rm il2} \\
        &- \Delta E_{\rm c}^{\rm il2/ch}
        + \Delta E_0^{\rm n} + \Delta E_{\rm F}^{\rm n} = 0,
\end{split}
\label{energy_SC}
\end{equation}
where $\Delta E_{\rm c}^{i/j} = E_{\rm c}^{i} - E_{\rm c}^{j}$ is the conduction band discontinuity between layers $i$ and $j$.
Using Eqs.~\eqref{ground state}-\eqref{fields_sc}, Eq.~\eqref{energy_SC} is rewritten as a closed form equation for $n_{\rm s}$:
\begin{equation}
    \begin{split}
         {\rm e}^{n_{\rm s}/n_{\rm cap}} \cdot
         {\rm e}^{n_{\rm s}/n_{\rm b}} \cdot
         &{\rm e}^{n_{\rm s}/n_{\rm il1}} \cdot
         {\rm e}^{n_{\rm s}/n_{\rm il2}} \cdot \\
         &{\rm e}^{(n_{\rm s}/n_{\rm 0})^{\frac23}} 
         ({\rm e}^{n_{\rm s}/N_{\rm c}^{\rm 2d}} - 1) 
         = 
         {\rm e}^{{-V_{\rm T}}/{V_{\rm th}}}.
     \end{split}
     \label{ns_SC}
\end{equation}
Here, $n_{i} = \frac{1}{q} C_{i} V_{\rm th}$, with $C_{i} = \tfrac{\epsilon_i}{t_i}$, and $V_{\rm th} = \frac{k_{\rm  b} T}{q}$.
$n_{\rm 0} = \frac{8}{9\pi} \frac{\epsilon_{\rm ch}}{q^2 \hbar} \sqrt{2m_{\rm c}^{*} (k_{\rm  b} T)^3}$ and $V_{\rm T}$ is the threshold voltage,
\begin{equation}
    \begin{split}
        V_{\rm T} = \Phi_{\rm b} - V_{\rm th}
        \Biggl(
        \frac{\sigma_\pi^{\rm cap/ch}}{n_{\rm cap}} &+
        \frac{\sigma_\pi^{\rm b/ch}}{n_{\rm b}} + \\
        \frac{\sigma_\pi^{\rm il1/ch}}{n_{\rm il1}} &+
        \frac{\sigma_\pi^{\rm il2/ch}}{n_{\rm il2}}
        \Biggr)
        + \frac{1}{q}\sum_{\rm j}^{\rm layers} \Delta E_{\rm c}^{j/j+1}.
    \end{split}
    \label{Vt_SC}
\end{equation}

\begin{figure}
    \centering
    \includegraphics[width=0.95\linewidth]{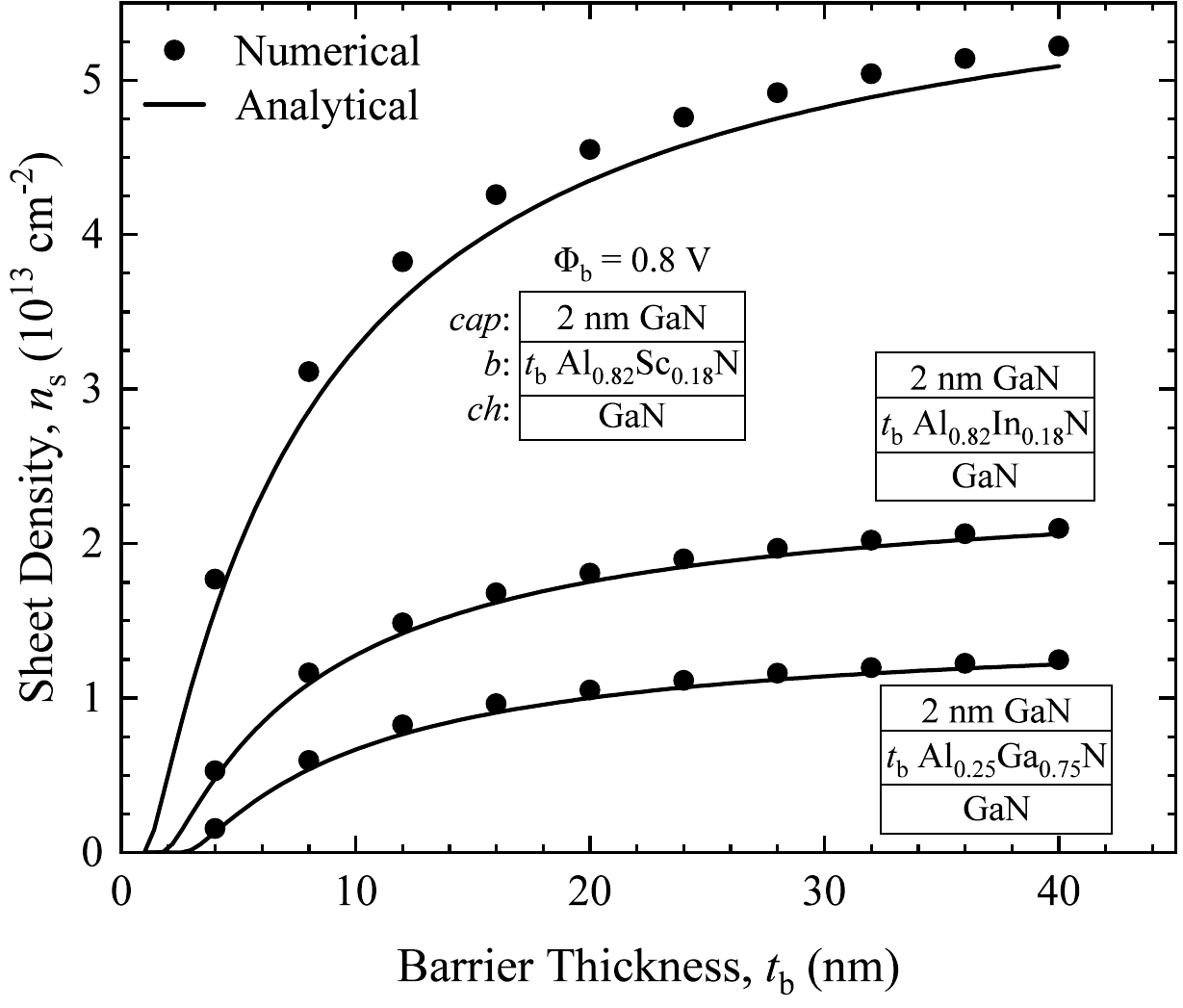}
    \caption{Calculated sheet density versus barrier thickness for single channel Al(Ga,In,Sc)N/ GaN heterostructures.
    A 2 nm GaN cap layer is considered, with a surface barrier height of 0.8 eV.
    No interlayers are considered.
    Solid lines correspond to results extracted from Eq.~\eqref{ns_SC}, while points are acquired from self-consistent numerical calculations using nextnano\cite{nextnano}.}
    \label{Fig2}
\end{figure}

The utility of Eq.~\eqref{ns_SC} is shown in Fig.~\ref{Fig2}, in which the 2DEG density is plotted against barrier thickness for AlScN/GaN, AlInN/GaN and AlGaN/GaN single channel heterostructures, capped by a 2 nm GaN layer.
The surface barrier height used is 0.8 V.
No interlayers were considered to clearly capture the 2DEG density dependence on barrier thickness.
The critical barrier thickness for the onset of 2DEG formation can be estimated using Eq.~\eqref{Vt_SC}, and demanding $V_{\rm T} = 0$, such that
\begin{equation}
\begin{split}
        t_{\rm b}^{\rm cr} =  \frac{\epsilon_{\rm b} V_{\rm th}}{q \sigma_\pi^{\rm b/ch}} 
        \Bigg[
        \frac{\Phi_{\rm b}}{V_{\rm th}} 
        - \Biggl(
        &\frac{\sigma_\pi^{\rm cap/ch}}{n_{\rm cap}} + 
        \frac{\sigma_\pi^{\rm il1/ch}}{n_{\rm il1}} + \\
        &\frac{\sigma_\pi^{\rm il2/ch}}{n_{\rm il2}}
        \Biggr)
        + \frac{1}{k_{\rm b} T} \sum_{\rm j}^{\rm layers} \Delta E_{\rm c}^{j/j+1}
        \Biggr].
\end{split}
\end{equation}
Said onset requirement is lowest for AlScN, increasing for AlInN, and highest for AlGaN for the alloy compositions considered, due to the polarization discontinuities between each barrier and the GaN channel.
As the barrier thickness increases, so does the expected 2DEG density, with larger polarization discontinuities inducing higher 2DEG densities.
Shown as points are results obtained from self-consistent numerical calculations using the Schr\"{o}dinger-Poisson solver nextnano\cite{nextnano}.
Numerical and analytical results are consistent, with a minor underestimation by the latter at high 2DEG densities due to the consideration of single subband occupancy by electrons. 
The material parameters used in our calculations are according to Ambacher et al.\cite{Ambacher2021}, and listed in Table \ref{Table1}.

The same approach will now be adopted to model the multichannel field effect transistor.

\begin{table}[ht!]
    \caption{Material parameters used in calculations.\cite{Ambacher2021}}
    \centering
    \begin{ruledtabular}
    \begin{tabular}{rcccccc}
        \multirow{ 2}{*}{Material} & $P_i$  & $\epsilon_i$ & $E_{\rm g}$ & $\Delta E_{\rm c}^{i/{\rm GaN}}$& $m_{\rm c}$ & $m_{\rm v}$ \\
         & (C/m$^2$) & ($\rm \epsilon_0$) & (eV) & (eV) & $ (m_{\rm e})$ & $(m_{\rm e})$ 
        \\
        \midrule
        GaN & 0.034 & 10.28 & 3.44 & - & 0.2 & 1.1 \\
        AlN & 0.148 & 10.31 & 6.16& 1.83 \\
        Al$_{\rm 0.25}$Ga$_{\rm 0.75}$N & 0.058 & 10.29 & 3.91 & 0.33 \\
        Al$_{\rm 0.82}$In$_{\rm 0.18}$N & 0.073 & 11.08 & 4.53 & 0.74 \\
        Al$_{\rm 0.82}$Sc$_{\rm 0.18}$N & 0.131 & 15.38 & 5.42 & 1.33\\
    \end{tabular}
    \end{ruledtabular}
    $m_{\rm e}$: free electron mass \hfill
    \label{Table1}
\end{table}

\section{Multi Channel} 

In order to model the multichannel field effect transistor, we make the assumption that the outer-most 2DEGs electrically isolate inner channels.
That is, the inner epitaxial periods are electrically independent from external effects, provided the weak nature of the latter compared to the band gap energies of employed materials.
The number of periods shall therefore bear no effect on the eventual energy landscape and mobile charge distribution.
Said electrical isolation manifests in the form of periodic boundary conditions across each inner epitaxial period.

\begin{figure*}[t!]
    \includegraphics[width=\textwidth]{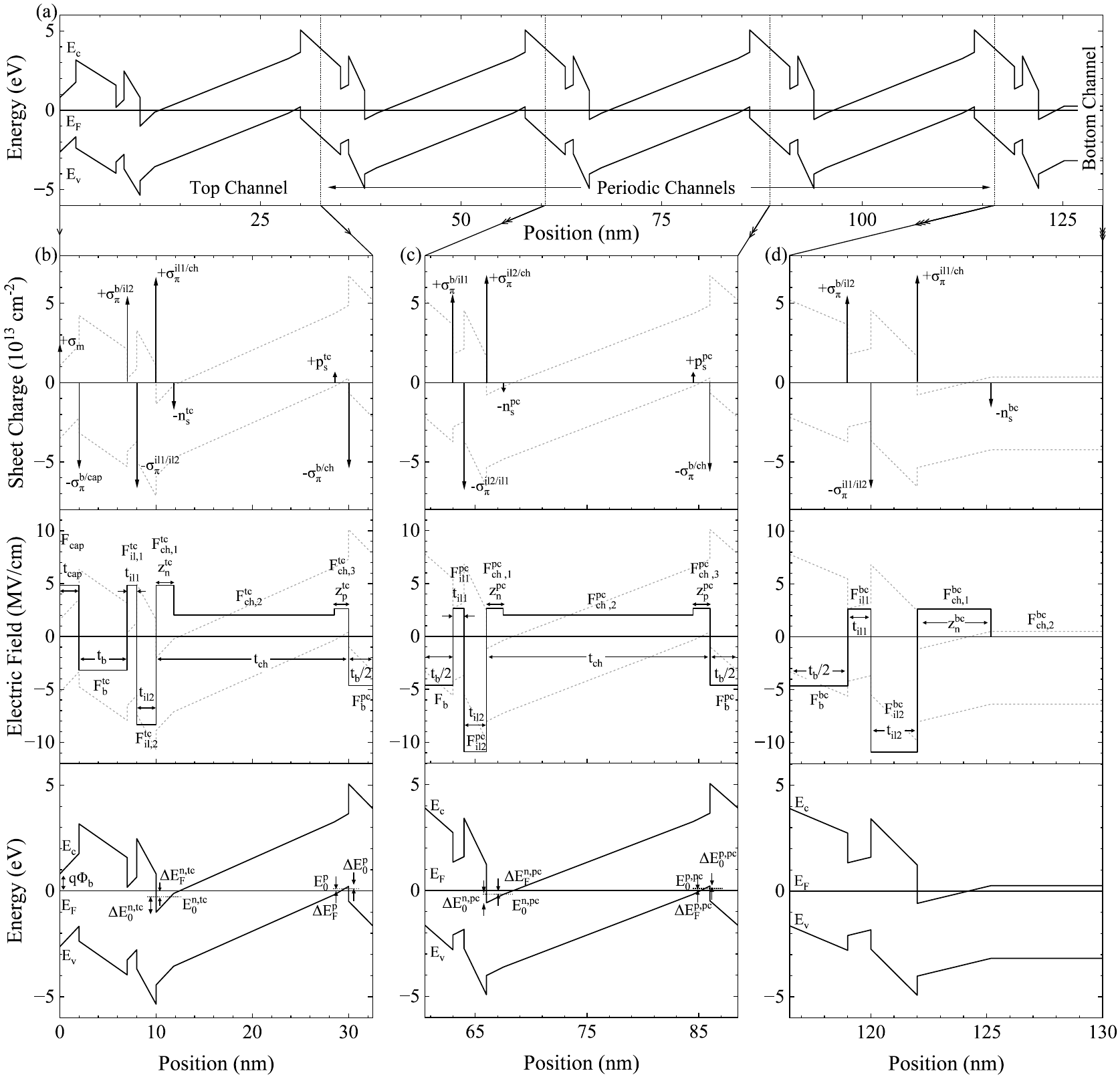}
    \caption{(a) Energy band diagram of an indicative five channel field effect transistor. The active region is divided into three regions of interest: the top channel, periodic channels, and the bottom channel. (b), (c), (d) Magnified charge, electric field, and energy band diagrams of the top, periodic and bottom channels, respectively. The energy band diagram is overlaid (dashed line) in the charge and field diagrams as a guide to the eye. All values shown are extracted from the analytical model. Notations are explained in the text.}
    \label{multichannel-big}
\end{figure*}

By this approach, the outline of this derivation is well defined.
We follow Fig.~\ref{multichannel-big}, and divide the structure into three regions; the top and bottom channels, separated by a sequence of periodic channels.
A closed form equation for the 2DEG/2DHG density is first derived for the periodic channels subject to periodic boundary conditions.
From there, the top and bottom channels are modeled by appropriately adjusting the respective boundary conditions to consistently align with the periodic stack on one end, and device termination on the other.
Hereinafter, we use the superscripts $tc$, $pc$ and $bc$ to distinguish between top, periodic and bottom channels.
Furthermore, the superscripts $n$ and $p$ will be used to refer to electron and hole magnitudes, respectively.

\subsection{Periodic Channels} \label{Periodic Channels}

Fig.~\ref{multichannel-big}(c) depicts the charge, electric field and energy band diagrams of a periodic undoped heterostructure.
Periodic boundary conditions imply the charge neutrality of each individual epitaxial period, which in turn dictates that -should the effect of background doping be negligible- the formation of a 2DEG ($n_{\rm s}^{\rm pc}$) is always accompanied by an adjacent 2DHG ($p_{\rm s}^{\rm pc}$) equal in density.
Equivalently,
\begin{equation}
   q n^{\rm pc}_{\rm s} = q p^{\rm pc}_{\rm s}.
\label{Charge_Neut_PC}
\end{equation}

Two relations can be extracted from the respective energy band diagram. 
The first follows the conduction band edge across the epitaxial period,
\begin{equation}
    \begin{split}
        qF^{\rm pc}_b t_{\rm b} -\Delta E_{\rm c}^{\rm b/il1}
        + qF^{\rm pc}_{\rm il1}t_{\rm il1} - \Delta &E_{\rm c}^{\rm il1/il2}  \\
        + qF^{\rm pc}_{\rm il2}t_{\rm il2} - \Delta E_{\rm c}^{\rm il2/ch} + qF^{\rm pc}_{\rm ch,1} &z_{\rm n} \\
        + qF^{\rm pc}_{\rm ch,2} (t_{\rm ch} - z_{\rm n} - z_{\rm p}) + qF^{\rm pc}_{\rm ch,3}& z_{\rm p} - \Delta E_{\rm c}^{\rm ch/b}= 0,
    \end{split}
    \label{energy1 periodic}
\end{equation}
while the second focuses on the channel region, such that
\begin{equation}
\begin{split}
        E_{\rm g}^{\rm ch} &+ \Delta E_0^{\rm n,pc} + \Delta E_{\rm F}^{\rm n,pc}  + \Delta E_{\rm F}^{\rm p,pc} + \Delta E_0^{\rm p,pc} = \\
        & qF^{\rm pc}_{\rm ch,1} z_{\rm n} + qF^{\rm pc}_{\rm ch,2} (t_{\rm ch} - z_{\rm n} - z_{\rm p}) + qF^{\rm pc}_{\rm ch,3} z_{\rm p},
\end{split}
\label{energy2 periodic}
\end{equation}
where $E_{\rm g}^{\rm ch}$ is the channel bandgap energy, and $z_{\rm n}$ and $z_{\rm p}$ denote the distance of the 2DEG and 2DHG from the respective interface.
Note that both electron and hole subbands appear. 
The positions of said subbands with respect to the bottom of the enclosing well are again approximated as
\begin{subequations}
\begin{eqnarray}
        \Delta E_0^{\rm n,pc} &&\approx \left( \frac{9\pi}{8}\right)^{\frac{2}{3}}  \Biggl[ \frac{(qF_{\rm ch,1}\hbar)^2}{2m_{\rm c}^{*}}\Biggr]^{\frac{1}{3}}, \\
        \Delta E_0^{\rm p,pc} &&\approx \left( \frac{9\pi}{8}\right)^{\frac{2}{3}}  \Biggl[ \frac{(qF_{\rm ch,3}\hbar)^2}{2m_{\rm v}^{*}}\Biggr]^{\frac{1}{3}},
    \end{eqnarray}
    \label{ground state periodic}
\end{subequations}
and
\begin{subequations}
\begin{eqnarray}
        \Delta E_{\rm F}^{\rm n} &&\equiv E_{\rm F} - E_{\rm 0}^{\rm n,pc} =  k_{\rm  b} T \ln \left({\rm e}^{n^{\rm pc}_{\rm s}/N_{\rm c}^{\rm 2d}} - 1 \right), \\
        \Delta E_{\rm F}^{\rm p} &&\equiv E_{\rm 0}^{\rm p,pc} - E_{\rm F} = k_{\rm  b} T \ln \left({\rm e}^{p^{\rm pc}_{\rm s}/N_{\rm v}^{\rm 2d}} - 1 \right).
    \end{eqnarray}
    \label{Fermi periodic}
\end{subequations}
In analogy to the single channel analysis, $m^{*}_{\rm v}$ is the hole effective mass, and $N_{\rm v}^{\rm 2d} = g_{\rm s} g_{\rm v}\frac{m^{*}_{\rm v}k_{\rm  b} T}{2\pi\hbar^2}$ is the valence band two-dimensional effective density of states.

We further note that the electric field terms within the channel are related by
\begin{subequations}
\begin{eqnarray}
    F^{\rm pc}_{\rm ch,2} &&= F^{\rm pc}_{\rm ch,1} - \frac{q n^{\rm pc}_{\rm s}}{\epsilon_{\rm ch}}, \label{Fch2_pc}\\
    F^{\rm pc}_{\rm ch,3} &&= F^{\rm pc}_{\rm ch,2} + \frac{q p^{\rm pc}_{\rm s}}{\epsilon_{\rm ch}} = F^{\rm pc}_{\rm ch,1},\label{Fch3_pc}
\end{eqnarray}
\end{subequations}
which allows us to solve for the electric field in the conduction band quantum well,

\begin{equation}
\begin{split}
        F^{\rm pc}_{\rm ch,1}
    = &\frac{V_{\rm th}}{t_{\rm ch}} 
    \Biggl(
    \alpha\frac{n^{\rm pc}_{\rm s}}{n_{\rm ch}} 
    + \frac{\sigma_\pi^{\rm b/ch}}{n_b} 
    + \frac{\sigma_\pi^{\rm il1/ch}}{n_{\rm il1}} 
    + \frac{\sigma_\pi^{\rm il2/ch}}{n_{\rm il2}}\\
     &+ \frac{1}{k_{\rm  b} T}\sum_{\rm j}^{\rm layers} \Delta E_{\rm c}^{j/j+1}
    \Biggr)
    \frac{
    \frac{1}{C_{\rm ch}}
    }{
    \frac{1}{C_{\rm b}} 
    + \frac{1}{C_{\rm il1}}
    + \frac{1}{C_{\rm il2}}
    +\frac{1}{C_{\rm ch}}
    }.
    \end{split}
    \label{F1}
\end{equation}
The parameter $\alpha \equiv \dfrac{t_{\rm ch} - z_{\rm n} - z_{\rm p}}{t_{\rm ch}} (\leq 1)$ captures the spread of the 2D carrier gases. 
Several sophisticated techniques can be employed to better estimate the position of each gas\cite{FangHoward}.
Alternatively, $\alpha$ can be treated as a fitting parameter to determine the distribution of the carrier gases.
For this work, we settle for the simplest approach whereby $\alpha$ is empirically treated and held constant.
It is typically expected to be close to unity, due to the narrow distribution of carriers ($\sim$ nm) compared to the much larger channel thickness ($\sim10$ nm).

Finally, the 2DEG density equation is acquired by substituting Eqs.~(\ref{ground state periodic})-(\ref{F1}) into Eq.~(\ref{energy2 periodic}),
\begin{equation}
    \begin{split}
    &({\rm e}^{n^{\rm pc}_{\rm s}/N_{\rm c}^{\rm 2d}}-1)
    ({\rm e}^{n^{\rm pc}_{\rm s}/N_{\rm v}^{\rm 2d}}-1) 
    \cdot
    {\rm e}^{\frac{
    \Delta E_0^{\rm n}(F^{\rm pc}_{\rm ch,1})}{k_{\rm  b} T}
    } 
    \cdot
    {\rm e}^{\frac{
    \Delta E_0^{\rm p}(F^{\rm pc}_{\rm ch,3})
    }{k_{\rm  b}T}} \\
    & \cdot \exp\left[\alpha {\frac{n_{\rm s}}{n_{\rm ch}}
    \frac{
    \frac{1}{C_{\rm b}} 
    + \frac{1}{C_{\rm il1}}
    + \frac{1}{C_{\rm il2}}
    }
    {
    \frac{1}{C_{\rm b}}
    +\frac{1}{C_{\rm il1}}
    +\frac{1}{C_{\rm il2}}
    +\frac{1}{C_{\rm ch}}}}\right]
    = {\rm e}^{-V_{\rm T}^{\rm pc}/V_{\rm th}},
        \end{split}
    \label{ns_equation}
\end{equation}
with 
\begin{equation}
\begin{split}
    V_{\rm T}^{\rm pc} = &\frac{1}{q}E_{\rm g}^{\rm ch} 
    - V_{\rm th} 
    \Biggl( 
    \frac{\sigma_\pi^{\rm b/ch}}{n_b}  
    + \frac{\sigma_\pi^{\rm il1/ch}}{n_{\rm il1}} 
    + \frac{\sigma_\pi^{\rm il2/ch}}{n_{\rm il2}} \\
    &+ \frac{1}{k_{\rm  b} T}\sum_{\rm j}^{\rm layers} \Delta E_{\rm c}^{j/j+1}
    \Biggr)
        \frac{\frac{1}{C_{\rm ch}}}{\frac{1}{C_{\rm b}}+ \frac{1}{C_{\rm il1}} + \frac{1}{C_{\rm il2}} + \frac{1}{C_{\rm ch}}}.
\end{split}
    \label{threshold_voltage}
\end{equation}

Eq.~(\ref{threshold_voltage}) corresponds to the critical condition for the formation of 2D carrier gases in periodic channels, allowing for the \textit{a priori} determination of 2DEG formation for any given epitaxial configuration. 
It can be intuitively understood by noticing that it can be written as,
\begin{equation}
    V_{\rm T}^{\rm pc} = \frac{1}{q}E_{\rm g}^{\rm ch} - F_{\rm ch,1}^{\rm pc}\biggr|_{n_{\rm s}^{\rm pc}=0} \cdot t_{\rm ch},
\end{equation}
where $F_{\rm ch,1}^{\rm pc}\biggr|_{n_{\rm s}^{\rm pc}=0}$ is the periodic electric field inside the channel induced by polarization charges alone.
If the potential drop across the channel is larger than its band gap energy, the Fermi level will cross into the conduction and valence bands, giving rise to mobile carrier gases.

\subsection{Top Channel}

The top channel derivation follows Fig.~\ref{multichannel-big}(b).
Starting from the top surface, and moving along the conduction band edge, we get

\begin{equation}
\begin{split}
        q \Phi_{\rm b} + q F_{\rm cap} t_{\rm cap} &- \Delta E_{\rm c}^{\rm cap/b}  + q F^{\rm tc}_{\rm b} t_{\rm b} \\ 
        - \Delta E_{\rm c}^{\rm b/il1} + q &F^{\rm tc}_{\rm il1} t_{\rm il1}  - \Delta E_{\rm c}^{\rm il1/il2} \\
         + q F^{\rm tc}_{\rm il2} t_{\rm il2} &- \Delta E_{\rm c}^{\rm il2/ch} + \Delta E_0^{\rm n,tc} + \Delta E_{\rm F}^{\rm n,tc} = 0.
\end{split}
\label{energy_tc}
\end{equation}

Moving from the underlying periodic channel to the top channel, we assume that no additional charge has been introduced, and the 2DHG density is rather unchanged.
Consequently, $F^{\rm tc}_{\rm ch,3} = F^{\rm pc}_{\rm ch,3}$ and $F^{\rm tc}_{\rm ch,2} = F^{\rm pc}_{\rm ch,2}$.
This allows us to write
\begin{equation}
    F^{\rm tc}_{\rm ch,1} = F^{\rm pc}_{\rm ch,2} + \frac{q n_{\rm s}^{\rm tc}}{\epsilon_{\rm ch}},
\end{equation}
which, similarly, leads to
\begin{equation}
\begin{split}
    {\rm e}^{n_{\rm s}/n_{\rm cap}} \cdot
    {\rm e}^{n_{\rm s}/n_{\rm b}} \cdot
    &{\rm e}^{n_{\rm s}/n_{\rm il1}} \cdot
    {\rm e}^{n_{\rm s}/n_{\rm il2}} \cdot \\
    & {\rm e}^{\frac{\Delta E_0^{\rm n,tc}(F^{\rm tc}_{\rm ch,1})}{k_{\rm  b} T}} \cdot
    ({\rm e}^{n^{\rm tc}_s/N_{\rm c}^{\rm 2d}}-1)
    = {\rm e}^{- V^{\rm tc}_{\rm T}/V_{\rm th}},
\end{split}
\label{ns_tc}
\end{equation}
with
\begin{equation}
\begin{split}
    V_{\rm T}^{\rm tc} &= \Phi_{\rm b} - 
    V_{\rm th}
    \Biggl(
    \frac{\sigma^{\rm cap/ch}_\pi}{n_{\rm cap}}+
    \frac{\sigma^{\rm b/ch}_\pi}{n_{\rm b}}+
    \frac{\sigma^{\rm il1/ch}_\pi}{n_{\rm il1}}+
    \frac{\sigma^{\rm il2/ch}_\pi}{n_{\rm il2}}
    \Biggr) \\
    +  
    &F^{\rm pc}_{\rm ch,2} t_{\rm ch}
    \Biggl(
    \frac{
    \frac{1}{C_{\rm ch}}
    }
    {
    \frac{1}{C_{\rm cap}} + 
    \frac{1}{C_{\rm b}} + 
    \frac{1}{C_{\rm il1}} + 
    \frac{1}{C_{\rm il2}}
    }
    \Biggr) ^{-1} 
    + \frac{1}{q}\sum_{\rm j}^{\rm layers} \Delta E_{\rm c}^{j/j+1} .
\end{split}
\label{Vt_tc}
\end{equation}
$F^{\rm pc}_{\rm ch,2}$ is calculated separately in Section \ref{Periodic Channels}.
Eqs.~\eqref{ns_tc} and \eqref{Vt_tc} are almost identical to those of the single channel, with the added information about the underlying layers through the aforementioned electric field term.

\subsection{Bottom Channel}
Modeling the bottom channel (Fig.~\ref{multichannel-big}(d)) is the simplest step in this derivation due to the electric field termination in the bulk substrate region.
By charge conservation, the electric field at the interface between the interlayer and the bottom channel, $F_{\rm ch,1}^{\rm bc}$ is the same as that of the equivalent periodic region. 
That is,
\begin{equation}
    F^{\rm bc}_{\rm ch,1} = F^{\rm pc}_{\rm ch,1}.
\end{equation}
Finally, requiring that the electric field vanishes in the bulk substrate region, the bottom channel 2DEG charge density must satisfy
\begin{equation}
    n_{\rm s}^{\rm bc} = \frac{1}{q} \epsilon_{\rm ch} F^{\rm pc}_{\rm ch,1}.
\end{equation}

Note that the triangular well approximation falls short in the  bottom-channel 2DEG density calculation under the previous assumptions.
Using Eqs.~\eqref{ground state} and \eqref{Fermi}, with $F^{\rm bc}_{\rm ch,1}$, would effectively repeat the periodic layout and fail to include information about device termination.
This could lead to a significant underestimation of mobile change density.
It can be further understood by considering the spatial distribution of the bottom channel 2DEG.
Said charge spread widens the respective quantum well, which in turn lowers the subband energy deeper into the well and increases its occupancy.

\subsection{Results}

Gathering the above results, the total mobile charge for any $N$-channel field effect transistor can be estimated by
\begin{subequations}
\begin{eqnarray}
    n_{\rm s}^{\rm tot} &&= n_{\rm s}^{\rm tc} + (N-2)\cdot n_{\rm s}^{\rm pc} + n_{\rm s}^{\rm bc}, \text{ and}\label{ns_tot}\\
    p_{\rm s}^{\rm tot} &&= (N-1)\cdot p_{\rm s}^{\rm pc}.\label{ps_tot}
\end{eqnarray}
\label{final result}
\end{subequations}
The derivation holds for the special case of $N=2$, wherein the top and bottom channels give rise to a single electrically isolated 2DHG whose density matches that of the periodic channels.

\begin{figure}[ht!]
    \centering
    \includegraphics[width=1\linewidth]{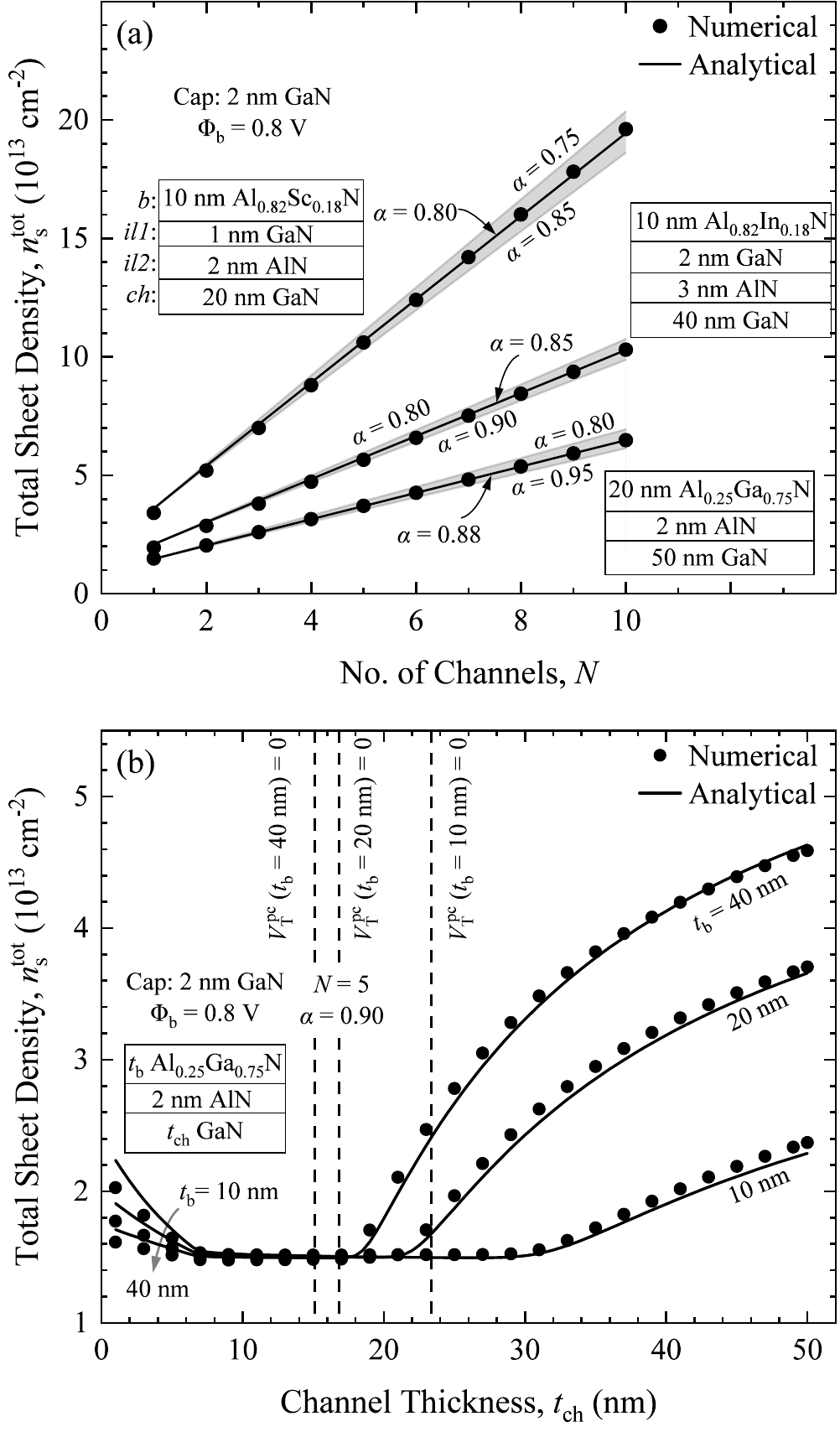}
    \caption{
    Total sheet density versus (a) the number of channels for Al(Ga,In,Sc)N/(GaN/AlN IL)/GaN heterostructures, and (b) channel thickness for a 5-channel AlGaN/2 nm AlN/GaN structure.
    A 2 nm GaN cap layer is considered, with a surface barrier height of 0.8 eV.
    Solid lines correspond to results obtained from Eqs.~\eqref{ns_SC} for $N=1$ and \eqref{final result} for $N\geq 2$, while points are acquired from self-consistent numerical calculations using nextnano\cite{nextnano}.
    The shaded areas in (a) correspond to ranges of $\alpha$ values as indicated. 
    }
    \label{Fig4}
\end{figure}

\begin{figure*}[ht!]
    \centering
    \includegraphics[width=1\textwidth]{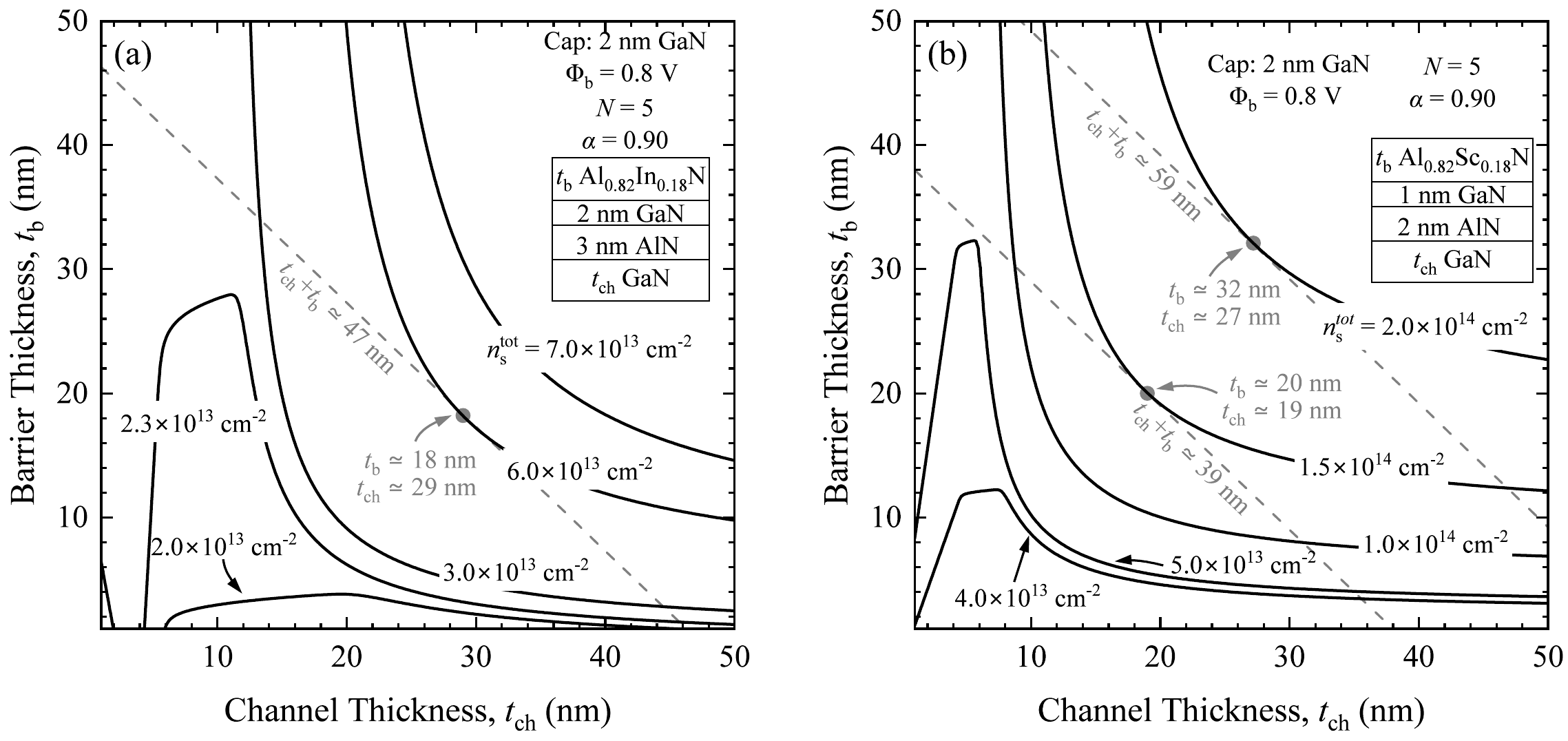}
    \caption{ Contour plots of total two-dimensional electron gas (2DEG) density as a function of channel and barrier thickness for 5-channel (a) AlInN/2 nm GaN/ 3 nm AlN/GaN and (b) AlScN/1 nm GaN/2 nm AlN/GaN heterostructures, respectively.
    A 2 nm GaN cap layer is considered, with a surface barrier height of 0.8 eV.
    A constant $\alpha$ value of 0.90 is used.
    The non-monotonic relation between 2DEG density and channel thickness stems from the 2DEG reduction in the bottom channel with channel thickness, which -prior to 2DEG formation within the periodic channels- constitutes the majority of mobile charge for this configuration.
    The dashed gray lines correspond to constant period thickness ($t_{\rm ch} + t_{\rm b}$) lines and indicate the thinnest period configuration necessary for the given total sheet carrier density.
    Results are obtained from Eq.~\eqref{final result}.}
    \label{Fig5}
\end{figure*}

Fig.~\ref{Fig4}(a) shows the total 2DEG density for GaN-based nitride heterostructures.
Al(Ga,In,Sc)N barriers are considered on GaN channels, with GaN/AlN interlayers, capped by a 2 nm GaN layer. 
The surface barrier height is assumed to be 0.8 V.
For AlGaN, no GaN interlayer is considered.
All layers are assumed fully strained on GaN.
We further assume that the conduction band discontinuities cancel each other, and the related sums therefore vanish.
Lines correspond to results acquired from the compact model developed in this work, and points are obtained from self-consistent numerical calculations using nextnano. 
The analytical results are consistent with numerical calculations.
For each case, a range of $\alpha$ values is used as indicated, which serves as an error margin for the above model.
When using more than two channels, the total charge density scales linearly with the increase rate corresponding to the periodic channel 2DEG density.

Said increase rate can be further perceived in Fig.~\ref{Fig4}(b), in which the total 2DEG density for a 5-channel AlGaN/2 nm AlN/GaN heterostructure as a function of channel and barrier thickness is shown.
For very small $t_{\rm ch}$, no 2DEG forms at the top or periodic channels ($V_{\rm T}^{\rm tc},~V_{\rm T}^{\rm pc} <0$), and the total sheet density consists of the bottom channel charge.
Note that the initial assumption regarding electrical isolation of inner channels becomes invalid when there is no mobile electron gas in the top channel, and this model's accuracy is diminished.
In this regime, increasing the channel thickness reduces the charge density.
As $t_{\rm ch}$ increases, the electric field within the channel region is lowered because $n_{\rm s}^{\rm pc}$ increases slowly compared to the reduction of $\frac{1}{t_{\rm ch}}$ in Eq.~\eqref{F1}. 
Consequently, the bottom channel is subject to a weaker electric field and accumulates fewer carriers.
At the same time, for the epitaxial configuration examined in Fig.~\ref{Fig4}(b), a very small channel thickness leads to electric field inversion in the AlGaN barrier. 
As a result, larger barrier thicknesses induce a stronger depletion of the bottom channel, reducing the total sheet density.
The vertical dashed lines indicate the critical channel thickness for the onset of 2DEG/2DHG formation within the periodic channels, or equivalently $V_{\rm T}^{\rm pc} = 0$.
Though not shown, the top channel is activated before the periodic channels.
Increasing the channel thickness above this boundary, mobile carriers accumulate within the remaining channels, and their contribution to the total density rapidly rises.
In this window, the same argument can be made for the barrier thickness, as thick barriers lower the channel thickness requirement the accumulation of mobile electrons.

The capacity of this work to facilitate experimental design is showcased in Fig.~\ref{Fig5}.
The latter is a contour plot of the total calculated 2DEG density for 5-channel AlScN/GaN and AlInN/GaN heterostructures as a function of channel and barrier thickness for $1~\mathrm{nm}\leq t_{\rm ch},t_{\rm b}\leq50$ nm.
Similar trends as previously noted are observed, with the total charge density rapidly scaling with the epitaxial thicknesses.
Dashed gray lines correspond to trajectories of constant period thickness, $t_{\rm ch} + t_{\rm b}$, as indicated.
When tangent to contour lines, the point of tangency is equal to the thinnest epitaxial configuration for the respective sheet density.
For example, if a total carrier density of $6\times10^{13}$ cm$^{-2}$ is desired, a viable option is a five-channel AlInN/2 nm GaN/3 nm AlN/ GaN heterostructure (Fig.~\ref{Fig5}a), with the thinnest stack comprising approximately 18 nm AlInN barriers and 29 nm GaN channels.
Inversely, for a given epitaxial scheme, the tangential contour corresponds to the highest sheet density possible.

Despite its accuracy, the model is subject to specific constraints. 
First, the parameter $\alpha$ must be carefully chosen.
Two-dimensional carrier gases typically span a region of the order of few nanometers.
As a result, for a given 2DEG/2DHG density, $z_{\rm n,p}$ become increasingly significant as $t_{\rm ch}$ is reduced, moving $\alpha$ to lower values.
This can be seen in Fig.~\ref{Fig4}(a), wherein thinner channels require lower values of $\alpha$ for analytical results to match numerical predictions.
Simultaneously, larger channel thicknesses result in 2DEGs of higher density, which inherently reside closer to the barrier/channel interface (assuming single subband occupation).
Therefore, as the 2DEG density increases, $z_{\rm n,p}$ are reduced, pushing $\alpha$ closer to unity. 
Second, the mobile charge spatial distribution is not accounted for.
As energy band discontinuities shrink, carrier wavefunctions spread into the barriers, which in turn affects the shape of each quantum well, and ultimately determines the energy landscape of the device.
In this work, said wavefunctions are considered to be exclusively within the channel layer, which limits the model's accuracy as the discontinuities shrink. 
Third, the derivation assumed the specific arrangement of mobile carriers discussed above.
Depending on the materials and thicknesses chosen for each layer, additional quantum wells might form, leading to carrier confinement outside the designated channel region.
To the same end, the band alignment between materials is assumed to be strictly straddling (type I).
Staggered (type II) or broken (type III) band alignment would significantly alter the presumed energy band diagram, and the consequent carrier confinement could drastically deviate from the above picture.

\section{Intentional Doping of Periodic Channels} \label{doping_section}

\begin{figure}[ht!]
    \centering
    \includegraphics[width=1\linewidth]{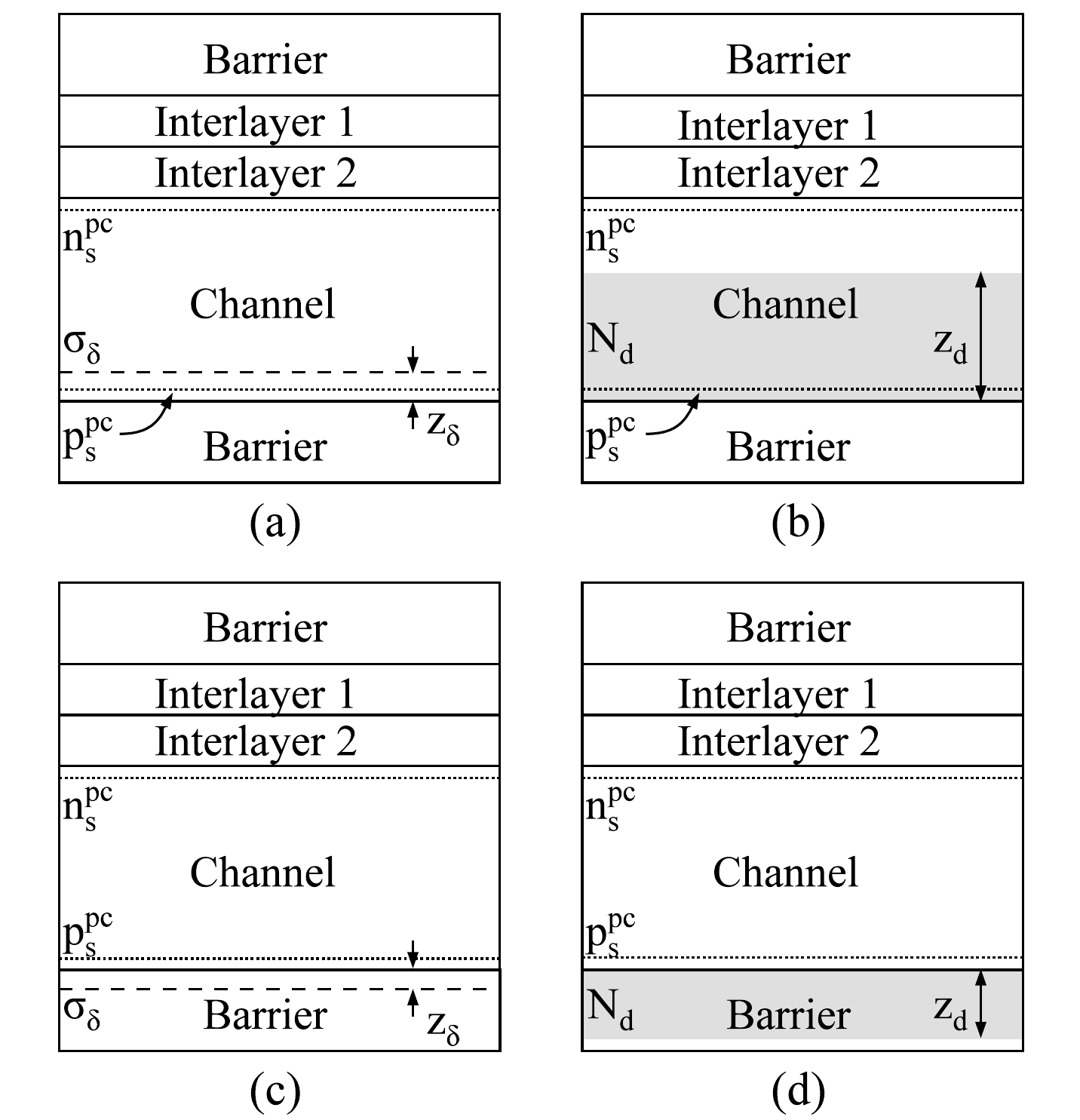}
    \caption{Cross-section schematics of donor doped periodic channels, with
    channel (a) $\delta$-doping, or (b) modulation doping, and barrier (c) $\delta$-doping, or (d) modulation doping.}
    \label{doping_schematic}
\end{figure}

The charge neutrality requirement above dictates that the formation of a 2DEG is always accompanied by a 2DHG (see Eq.~(\ref{Charge_Neut_PC})).
For multichannel transistors, however, undesirable 2DHGs, constitute parasitic parallel conduction channels which deteriorate device performance.
In this section we investigate the efficacy of hole gas depletion techniques by intentional compensation by either delta, or modulation doping with donors, following Fig.~\ref{doping_schematic}. 

The delta doping density is denoted by $\sigma_\delta$ (in cm$^{-2}$) , and $z_\delta$ is its position, while the modulation doping concentration is $N_{\rm d}$ (in cm$^{-3}$). 
The use of a spacer layer between doping regions and the conductive layer is common practice for mobility enhancement.
To this end, we consider modulation doping over only a region $z_{\rm d}$ of the enclosing layer.
Notice the different conventions used for the distances $z_\delta$ and $z_{\rm d}$.
For channel doping, they are measured with respect to the bottom of the doped layer, and with respect to the top of the layer for barrier doping. 
Dopants are assumed to be fully ionized. 

\begin{figure*}[ht!]
    \includegraphics[width=1\textwidth]{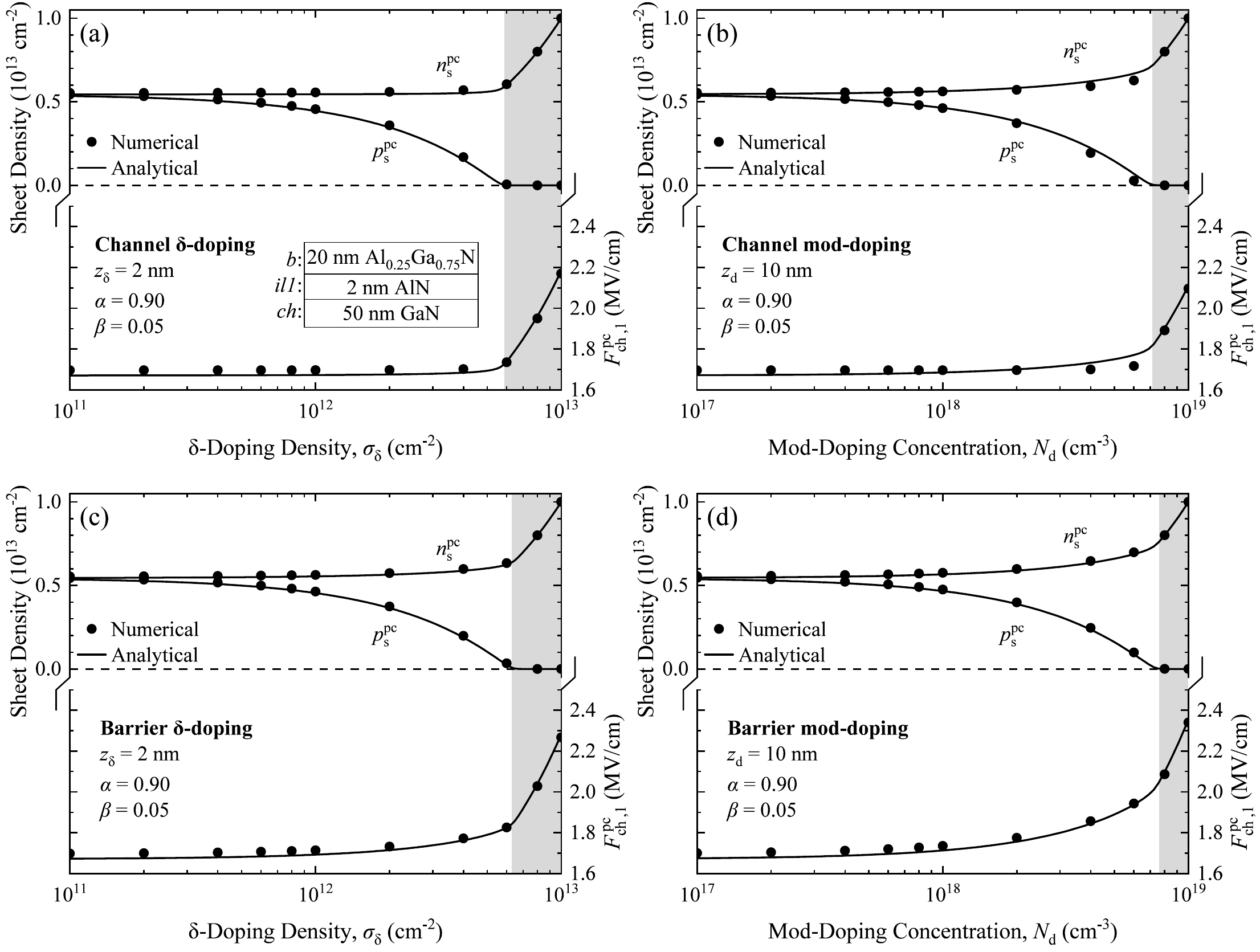}
    \caption{Calculated sheet charge density, and electron quantum well electric field in AlGaN/AlN/GaN periodic channels versus doping density for (a) channel $\delta$-doping, (b) channel modulation doping, (c) barrier $\delta$-doping, and (d) barrier modulation doping.
    Solid lines correspond to results extracted from Eqs.~\eqref{ns_doping} and \eqref{F1_doping}, and points are acquired from self-consistent numerical calculations using nextnano\cite{nextnano}.
    The shaded areas indicate the complete depletion of mobile holes.
    }
    \label{Fig7}
\end{figure*}

The approach remains the same.
The main difference appears in the charge neutrality equation, which now becomes
\begin{subequations}
    \begin{eqnarray}
        q n_{\rm s} = q p_{\rm s} + q\sigma_\delta,&&~~\text{for $\delta$-doping, and} \label{dd-charge-neut}\\
        q n_{\rm s} = q p_{\rm s} + qN_{\rm d} \cdot z_{\rm d},&&~~\text{for modulation doping.}   \label{md-charge-neut}
    \end{eqnarray}\label{doped_charge_neut}
\end{subequations}
Eqs.~\eqref{doped_charge_neut} imply that donor-generated charge will strictly exist within the periodic channels, either compensating the 2DHG or enhancing the 2DEG.
In other words, we assume that donor doping bears no direct effect on the outer most channels, but only indirectly does so by altering the respective boundary conditions.

The final periodic-channel 2DEG density equation is of the form
\begin{equation}
    \begin{split}
    ({\rm e}&^{n^{\rm pc}_{\rm s}/N_{\rm c}^{\rm 2d}}-1)
    ({\rm e}^{(n^{\rm pc}_{\rm s}-\sigma_{\rm q})/N_{\rm v}^{\rm 2d}}-1) 
    \cdot
    {\rm e}^{\frac{
    \Delta E_0^{\rm n,pc}(F^{\rm pc}_{\rm ch,1})}{k_{\rm  b} T}
    }
    \\
    \cdot
    &{\rm e}^{\frac{
    \Delta E_0^{\rm p,pc}(F^{\rm pc}_{\rm ch,3})
    }{k_{\rm  b}T}}
    \cdot 
    \exp
    \left[ 
    \left(
    \alpha \frac{n^{\rm pc}_{\rm s}}{n_{\rm ch}}
    + \beta \frac{\sigma_{\rm q}}{n_{\rm ch}}
    \right)
    \cdot
    \Tilde{C}_1
    \right]
    = {\rm e}^{-V^{\rm pc}_{\rm T}/V_{\rm th}},
        \end{split}
    \label{ns_doping}
\end{equation}

\noindent where 
$\sigma_{\rm q} = \sigma_\delta$ or $\sigma_{\rm q} = N_{\rm d}\cdot z_{\rm d}$ determined by the doping scheme,
and $\Tilde{C}_1 = \frac{\frac{1}{C_{\rm b}} + \frac{1}{C_{\rm il1}}+ \frac{1}{C_{\rm il2}}}{\frac{1}{C_{\rm b}}+\frac{1}{C_{\rm il1}}+ \frac{1}{C_{\rm il2}}+\frac{1}{C_{\rm ch}}}$. 
The parameter $\beta \equiv \tfrac{z^{\rm pc}_{\rm p}}{t_{\rm ch}}$ captures the 2DHG position and, like $\alpha$, is treated empirically.
The electric field in the conduction band quantum well, $F^{\rm pc}_{\rm ch,1}$, and the critical voltage, $V^{\rm pc}_{\rm T}$, follow a similar behavior, with
\begin{equation}
\begin{split}
    F^{\rm pc}_{\rm ch,1}
    = \frac{V_{\rm th}}{t_{\rm ch}} 
    \Biggl( 
    &\alpha\frac{n^{\rm pc}_{\rm s}}{n_{\rm ch}} 
    + \beta\frac{\sigma_{\rm q}}{n_{\rm ch}} 
    + \frac{\sigma_\pi^{\rm b/ch}}{n_b} 
    + \frac{\sigma_\pi^{\rm il1/ch}}{n_{\rm il1}}
    \\+ \frac{\sigma_\pi^{\rm il2/ch}}{n_{\rm il2}}
    &+ \frac{1}{k_{\rm  b} T}\sum_{\rm j}^{\rm layers} \Delta E_{\rm c}^{j/j+1}
    \Biggr)
    \cdot
    \Tilde{C}_2
    + \Delta F_{\rm q},
        \label{F1_doping}
\end{split}
\end{equation}
and
\begin{equation}
\begin{split}
V^{\rm pc}_{\rm T} = \frac{1}{q}E_{\rm g}^{\rm ch} 
    - V_{\rm th}  
    \Biggl( 
    \frac{\sigma_\pi^{\rm b/ch}}{n_b} 
    &+ \frac{\sigma_\pi^{\rm il1/ch}}{n_{\rm il1}}
    + \frac{\sigma_\pi^{\rm il2/ch}}{n_{\rm il2}} 
    \Biggr)
    \cdot
    \Tilde{C}_2
     \\
     &+ \frac{1}{q}\sum_{\rm j}^{\rm layers} \Delta E_{\rm c}^{j/j+1}
     - \Delta V_{\rm q},
    \label{Vt-doping}
    \end{split}
\end{equation}
where $\Tilde{C}_2 = 1 - \Tilde{C}_1 = \frac{\frac{1}{C_{\rm ch}}}{\frac{1}{C_{\rm b}} + \frac{1}{C_{\rm il1}} + \frac{1}{C_{\rm il2}}+ \frac{1}{C_{\rm ch}}}$.
The terms $\Delta F_{\rm q}$ and $\Delta V_{\rm q}$ are design parameters, controlled by the doping configuration.
The results are summarized in Table \ref{generalized_equation}.
This generalized form further covers the undoped structure, for which all additional terms vanish by setting $\sigma_{\rm q} = 0$. 

\begin{table*}[tp]
    \centering
    
    \begin{ruledtabular}  
    \begin{tabular*}{\textwidth}{@{}r@{\hspace{3em}}c@{\hspace{4em}}c@{\hspace{3em}}c@{}c@{}}
    
        2DEG Density Equation &   
    \multicolumn{4}{@{}c}{
    $
    ({\rm e}^{n^{\rm pc}_{\rm s}/N_{\rm c}^{\rm 2d}} - 1)
    ({\rm e}^{(n^{\rm pc}_{\rm s}-\sigma_{\rm q})/N_{\rm v}^{\rm 2d}} - 1) 
    \cdot
    {\exp}\left({\dfrac{
    \Delta E_0^{\rm n,pc}(F_{\rm ch,1}^{\rm pc})}{k_{\rm  b} T}
    }\right) 
    \cdot
    {\exp}\left({\dfrac{
    \Delta E_0^{\rm p,pc}(F_{\rm ch,3}^{\rm pc})
    }{k_{\rm  b}T}}\right)\cdot$ }
    \\ 
     & & &
    \multicolumn{2}{@{}c}{
    $\cdot \exp\left[\left( \alpha\dfrac{n^{\rm pc}_{\rm s}}{n_{\rm ch}} +\beta\dfrac{\sigma_{\rm q}}{n_{\rm ch}}\right) \cdot 
    \Tilde{C}_1
    \right]
    = {\rm e}^{-V^{\rm pc}_{\rm T} / V_{\rm th}}
    $ 
} 
    \\
    \\
    Electric Field in Electron Quantum Well & \multicolumn{4}{@{}l}{
    $
    F^{\rm pc}_{\rm ch,1}
    = \dfrac{V_{\rm th}}{t_{\rm ch}} 
    \left(
    \alpha\dfrac{n^{\rm pc}_{\rm s}}{n_{\rm ch}} + 
    \beta \dfrac{\sigma_{\rm q}}{n_{\rm ch}} 
    + \dfrac{\sigma_\pi^{\rm b/ch}}{n_b} 
    + \dfrac{\sigma_\pi^{\rm il1/ch}}{n_{\rm il1}}
    \right.
    $
    }
    \\
    & & & \multicolumn{2}{@{}c}{
    $
    \left.
    + \dfrac{\sigma_\pi^{\rm il2/ch}}{n_{\rm il2}}
    + \dfrac{1}{k_b T}\sum_{\rm j}^{} \Delta E_{\rm c}^{j/j+1}
    \right)
    \cdot
    \Tilde{C}_2
    + \Delta F_{\rm q}
    $
    }
    \\
    \\
    Critical Condition for 2DEG Formation & \multicolumn{4}{@{}c}{
    $
    V^{\rm pc}_{\rm T} = \dfrac{1}{q}E_{\rm g}^{\rm ch}
    - V_{\rm th} 
    \left( 
    \dfrac{\sigma_\pi^{\rm b/ch} }{n_b} 
    + \dfrac{\sigma_\pi^{\rm il1/ch} }{n_{\rm il1}} 
    + \dfrac{\sigma_\pi^{\rm il2/ch}}{n_{\rm il2}}
    \right)
    \cdot
    \Tilde{C}_2
    + \dfrac{1}{q}\sum_{\rm j}^{} \Delta E_{\rm c}^{j/j+1}
    -\Delta V_{\rm q}
    $
    }
    \\
    \midrule\midrule
    Doping Scheme 
    & $\sigma_{\rm q}$
    & $F^{\rm pc}_{\rm ch,3}$
    & $\Delta F_{\rm q}$ 
    & $\Delta V_{\rm q}$     
    \\
    \midrule
    Undoped 
    & 0
    & $F^{\rm pc}_{\rm ch,1}$
    & 0 
    & 0 
    \\ \\
    Channel $\delta$-doping        
    & $\sigma_\delta$
    &  $F^{\rm pc}_{\rm ch,1}$
    & $-\dfrac{V_{\rm th}}{t_{\rm ch}}\dfrac{z_\delta}{t_{\rm ch}} \dfrac{\sigma_{\rm q}}{n_{\rm ch}} \Tilde{C}_2$ 
    & $V_{\rm th} \dfrac{z_\delta}{t_{\rm ch}} \dfrac{\sigma_{\rm q}}{n_{\rm ch}}  \Tilde{C}_1$
    \\ \\
    Barrier $\delta$-doping        
    & $\sigma_\delta$
    & $F^{\rm pc}_{\rm ch,1} - \dfrac{q\sigma_{\rm q}}{\epsilon_{\rm ch}}$
    & $\dfrac{V_{\rm th}}{t_{\rm ch}} \dfrac{z_\delta}{t_{\rm b}} \dfrac{\sigma_{\rm q}}{n_{\rm b}} \Tilde{C}_2$ 
    & $ V_{\rm th} \dfrac{z_\delta}{t_{\rm b}} \dfrac{\sigma_{\rm q}}{n_{\rm b}} \Tilde{C}_2$
       
    \\ \\
    Channel modulation-doping     
    & $N_{\rm d}\cdot z_{\rm d}$
    & $F^{\rm pc}_{\rm ch,1}$  
    & $-\dfrac{1}{2}\dfrac{V_{\rm th}}{t_{\rm ch}} \dfrac{z_{\rm d}}{t_{\rm ch}} \dfrac{\sigma_{\rm q}}{n_{\rm ch}} \Tilde{C}_2$
    &   $\dfrac{1}{2}V_{\rm th} \dfrac{z_{\rm d}}{t_{\rm ch}} \dfrac{\sigma_{\rm q}}{n_{\rm ch}} \Tilde{C}_1$

    \\ \\
    Barrier modulation-doping     
    & $N_{\rm d}\cdot z_{\rm d}$
    & $F_{\rm ch,1}^{\rm pc} - \dfrac{q\sigma_{\rm q}}{\epsilon_{\rm ch}}$ 
    & $\dfrac{1}{2}\dfrac{V_{\rm th}}{t_{\rm ch}} \dfrac{z_{\rm d}}{t_{\rm b}} \dfrac{\sigma_{\rm q}}{n_{\rm b}} \Tilde{C}_2$
    &  $\dfrac{1}{2}V_{\rm th} \dfrac{z_{\rm d}}{t_{\rm b}} \dfrac{\sigma_{\rm q}}{n_{\rm b}} \Tilde{C}_2$  
    \\
    \midrule
      \multicolumn{4}{@{}l}{$\Tilde{C}_1 = \dfrac{\frac{1}{C_{\rm b}}+\frac{1}{C_{\rm il1}}+\frac{1}{C_{\rm il2}}}{\frac{1}{C_{\rm b}} + \frac{1}{C_{\rm il1}} + \frac{1}{C_{\rm il2}} + \frac{1}{C_{\rm ch}}},\ \  \Tilde{C}_2 = 1 - \Tilde{C}_1= \dfrac{\frac{1}{C_{\rm ch}}}{\frac{1}{C_{\rm b}} + \frac{1}{C_{\rm il1}} + \frac{1}{C_{\rm il2}} + \frac{1}{C_{\rm ch}}} $
      }    
    \end{tabular*}
    \caption{Generalized 2DEG density equation for periodic channels of a MCFET for the various doping schemes examined in this work.}
        \label{generalized_equation}
    \end{ruledtabular}
\end{table*}

Fig.~\ref{Fig7} shows the calculated 2DEG and 2DHG densities, and the electron quantum well electric field across the doping schemes examined for periodic channels in an 20 nm AlGaN/2 nm AlN/50 nm GaN heterostructure, using Eqs.~\eqref{ns_doping} and \eqref{F1_doping}.
For $\delta$-doping, we use $z_{\rm \delta} = 2$ nm, and for modulation doping, $z_{\rm d} = 10$ nm.
Across all configurations, $\alpha = 0.90$ and $\beta = 0.05$ are used.
The modulation doping concentrations considered are such that the donor charge introduced is the same for all doping schemes.
That is, $\sigma_{\rm q} = \sigma_\delta = N_{\rm d} \cdot z_{\rm d} \in [10^{11}, 10^{13}]$ cm$^{-2}$.
For said reason, the efficacy of 2DHG depletion for each doping technique is similar.
The compact model is consistent with numerical calculations.

\begin{figure*}[ht!]
    \centering
    \includegraphics[width=1\textwidth]{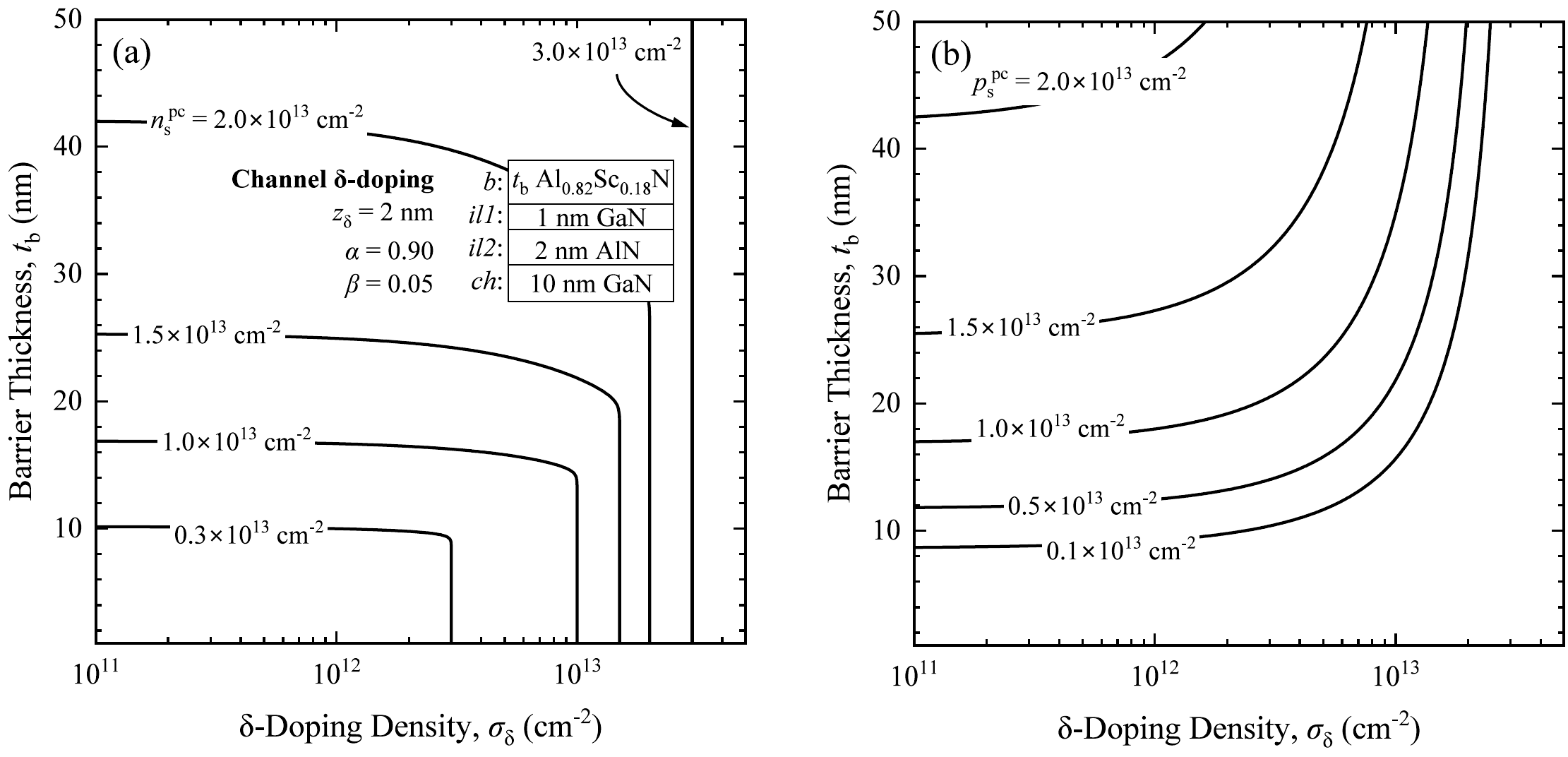}
    \caption{Calculated two-dimensional (a) electron and (b) hole gas density within an individual AlScN/1 nm GaN/2 nm AlN/10 nm GaN periodic channel as a function of barrier thickness and channel $\delta$-doping density.
    The doping plane is 2 nm above the bottom GaN/AlScN interface. 
    Constant $\alpha$ and $\beta$ values of 0.90 and 0.05 are used.
    Results are obtained from Eq.~\eqref{ns_doping}.
    }
    \label{Fig8}
\end{figure*}

Another benefit that comes with doping is the miniaturization of the epitaxial stack.\cite{Sohi_2021}
By charge neutrality, donor-doping provides precise charge control over the 2DEG density, allowing for considerable device thinning without the sacrifice of mobile carriers, as seen in Fig.~\ref{Fig8}.
The latter shows the 2DEG (Fig.~\ref{Fig8}a) and 2DHG (Fig.~\ref{Fig8}b) densities within an individual periodic channel.
Sufficiently high donor doping densities compensate the mobile hole gas, and further render the 2DEG density independent of barrier thickness.
From a practical perspective, this considerably simplifies epitaxy, while the thinner epitaxial stack facilitates device fabrication.

However, special attention must be paid when choosing a doping configuration.
Pushing for very high doping concentrations can potentially lead to the formation of an additional quantum well near the doping region, giving rise to an undesirable parasitic channel.
This phenomenon is not captured in our derivation and poses an additional limitation of the compact model developed in this work.
The second-well formation threshold can be estimated for any configuration by seeking the $\sigma_{\rm q}$ value that sets $F_{\rm ch,2}^{\rm pc}$ to zero.
Doping densities above that would push $F_{\rm ch,2}^{\rm pc}$ to negative values, giving rise to the parasitic electron channel. 

Beside mobile carrier density, doping can further control carrier mobility.
Interface roughness scattering in quantum wells is strongly dependent on quantum well width and the electric field therein.\cite{Jana2011}
By Table \ref{generalized_equation}, doping affects the electric field in the conduction band quantum well.
Following Fig.~\ref{Fig7}, the electron quantum well electric field remains rather constant while the 2DHG is compensated by donors.
Once the 2DHG is completely depleted, excess doping will instead enhance the existing 2DEG, and significantly strengthen the respective electric field.
A minor difference in electric field strength is seen between the doping schemes, with channel doping exhibiting slightly weaker electric fields.
In parallel, remote Coulomb scattering can further damage carrier mobility, and therefore doping must be carefully engineered to optimize conduction.

\section{Conclusions}

A system of closed-form equations has been derived for the total mobile carrier density in multichannel polar heterostructures.
Multilayered heterostructures based on the III-Nitride material system are widely employed in electronic and photonic applications.
Their increased popularity gives rise to an imperative need for an accurate, efficient and transparent design framework.
The generalized analytical model presented in this work can be used for the determination of mobile carrier gas densities in such devices, and allows for the analytic investigation of each design parameter, with limited empiricism, across multiple doping configurations.
Notwithstanding the specific cases examined above, this work sets the ground floor for the modeling of any periodic structural design regardless of complexity.
It provides an efficient path for experiment design that complements numerical simulations and paves the way for the development of next-generation III-Nitride semiconductor devices.
It finally lays the foundation for the analysis of source, drain, and gate characteristics, which will be the focus of the next stage of this study to complete the compact model for multichannel field effect transistors.

\section*{Acknowledgments}
This work was supported in part by SUPREME, one of seven centers in JUMP 2.0, a Semiconductor Research Corporation (SRC) program sponsored by DARPA. This prototype (or technology) was partially supported by the Microelectronics Commons Program, a DoW initiative, under award number N00164-23-9-G061.

\section*{Author Declarations}
\subsection*{Conflict of Interest}
The authors have no conflicts to disclose.
\section*{Data Availability}

The data that supports the findings of this study are available within the article and its supplementary material.

\section*{References}
\bibliography{CCM_ref}

\end{document}